\documentclass{ws-procs9x6}
\usepackage{graphics} 
\def\be{\begin{eqnarray}&&} \def\ee{\end{eqnarray}}
\def \nonu {\nonumber \\&&}

\newcommand{\pol}[1] {\stackrel{\rightarrow}{#1}}
\begin{document}

\title{A Study of Final State Effects  \\ in the electrodisintegration 
 of a polarized Helium-3 target
}
\author{A. Kievsky}

\address{Istituto Nazionale di Fisica Nucleare, 
Sezione di Pisa, Via
Buonarroti 2, I-56100
Pisa, Italy}

\author{E. Pace}

\address{Dipartimento di Fisica, Universit\`{a} di Roma "Tor Vergata" 
and
Istituto Nazionale di Fisica Nucleare, Sezione Tor Vergata, Via della Ricerca
Scientifica 1, I-00133 Roma, Italy}

\author{G. Salm\`e}

\address{Istituto Nazionale di Fisica Nucleare, 
Sezione Roma I, P.le A. Moro
2, I-00185 Roma, Italy}  

\maketitle

\abstracts{ An approach for the description of the
 final state interaction in the evaluation of inclusive electromagnetic responses 
of a polarized
$^3$He target, is briefly illustrated. Preliminary results of calculations, 
where  
the final state interaction is fully
taken into account for the two-body break-up channel,  are compared  
 with experimental data, showing a very encouraging improvement with respect to the plane
wave impulse approximation results.  
}

\section{Introduction}
 
  The relevance of polarized $^3$He targets for investigating the electromagnetic (em)
  properties of the neutron is well known, and a huge amount of experimental and
  theoretical  work has been devoted to this  issue (see, e.g., Refs.
  \refcite{Xu}, \refcite{Xu2} and 
  \refcite{a1n} for an overview of the experimental status).    
The main motivations for such a viable activity is the absence in nature
of free-neutron targets and the attempt of devising an effective neutron target,
by means of a polarized $^3$He target. From the theoretical side, 
the difficulties come from  the non  trivial task
of  disentangling 
the neutron
information from the nuclear-structure effects. In particular many
effects may play a relevant role, like: i) the "small" components of the bound-state wave
function; ii) the $\Delta$ excitation; iii) 
the inclusion of the final state interaction
(FSI), between the knocked-out nucleon and the interacting spectator pair; 
iv) the
meson exchange currents (MEC); v) relativity.
As a first step, the analysis of inclusive em responses of  polarized $^{3}$He, 
in the region of the quasi-elastic (qe) peak, was carried out    within the 
{\em plane wave impulse approximation} 
(PWIA)\cite{CPS92,CPS95,KPSV,Sauer}, where realistic 
nucleon-nucleon interactions and a relativistic electron-nucleon
 cross section were adopted. In the final
state, only the interaction between the interacting spectator pair 
and the knocked-out
nucleon was disregarded.

Recently, a step forward was performed through  calculations 
of the em responses which 
include FSI, but within a  non relativistic framework\cite{Glock1,Glock2}.

In this contribution, a  preliminary report (see also Ref. \refcite{KPS04}) of   the inclusive
em responses  calculated by taking into account both relativistic effects and 
 FSI in the two-body break-up 
channel, will be presented.
 
\section{The polarized cross section in PWIA}   
The inclusive scattering of polarized electrons (with helicity $h$) by a polarized $^3$He target
($ \stackrel{\rightarrow}{e}~ +~^3\! \pol{\rm  He}~ \rightarrow ~e'~ +~X$)
is given by
\be {d^2\sigma(h) \over d\Omega d\omega}~=~
\Sigma\;+\;h\;\Delta \label{xsect} \ee
with
\be {\Sigma}~=~ \sigma_{Mott}~ \left [
 \left({Q^2 \over |\vec{q}|^2}\right
 )^2~R_{L}(Q^2,\omega) + \left({Q^2 \over
 2|\vec{q}|^2}+tan^2{\theta_{e}\over 2}\right)~R_{T}(Q^2,\omega)\vphantom{\left({Q^2 \over |\vec{q}|^2}\right
 )^2}\right]
 \label{Sigma} \ee 
 \be \Delta~=~
 -\sigma_{Mott}~tan\frac{\theta_{e}}{2}~
 \left\{cos\theta^*~R_{T'}(Q^2,\omega)~
 {\left(\epsilon_i+\epsilon_f \right) \over |\vec{q}|} 
 tan\frac{\theta_{e}}{2}~+ \right.  \nonu \left.
 -~{Q^2 \over |\vec{q}|^2~\sqrt{2}}~sin\theta^*
 cos\phi^*~R_{TL'}(Q^2,\omega)\right \} \label{Delta} \ee
where $\theta_{e}$ is the scattering angle, $\epsilon_{i(f)}$  the energy of the
initial (final) electron, $\theta^* $ and $\phi^*$
 are the azimuthal and polar angles of the target polarization vector, 
 with respect to the direction of the three-momentum
 transfer $\vec{q}$; $Q^2=|\vec{q}|^2 - \omega^2$, with $\omega$ the energy transfer. 
 The unpolarized ($R_{L}$,
$R_{T}$) and polarized ($R_{T'}$, $R_{TL'}$) inclusive responses contain the
nuclear-structure effects. Finally the asymmetry, A, is defined as 
A$=\Delta / \Sigma$.

 In order to fully appreciate the step forward represented by fully including 
 the interaction  in the final three-nucleon system,
 let us remind that in  our PWIA calculations of the inclusive em responses 
 the following approximation for  
   the final three-nucleon system
is considered 
\be |j,j_z,T,T_z,\pi,\epsilon_{int},\alpha;{\bf P} \rangle  \rightarrow 
 {1 \over \sqrt{3}} 
|{\bf p}_f,\sigma_f \tau_ f \rangle  |\alpha, m _{23},
 \tau_{23};{\bf P}_{23} \rangle \label{PWIA}\ee
with $j(j_z)$   
the total angular momentum (third component), $T(T_z)$ the total isospin (third
component), $\pi$ the total parity, $\epsilon_{int}$ the
intrinsic energy of the  three-nucleon system, 
$\alpha \equiv \{j_{23} T_{23} \lambda_{23}, \pi _{23},
\epsilon_{23}\}$,
${\bf P}= {\bf p}_f +{\bf P}_{23} $ the three-momentum of the three-nucleon center of
mass (CM), 
  $|{\bf p}_f,\sigma_f \tau_ f \rangle$   the plane wave
describing the knocked-out nucleon. In Eq. (\ref{PWIA}), the wave function of the fully-interacting 
spectator pair is given by $|\alpha, m_{23},
 \tau_{23};{\bf P}_{23} \rangle$.
The  terms that 
properly antisymmetrize the approximated three-nucleon wave function, are dropped
out, 
since only the direct
interaction between the virtual photon and the struck nucleon is taken into
account. 

In Fig. 1,
the experimental asymmetry A (proportional to  the transverse (polarized) response, 
$R_{T'}$, at the qe peak) recently measured at TJLAB,  in the region of the
qe peak\cite{Xu}, is compared to our PWIA calculations\cite{KPSV}, 
obtained  by using the AV18 nucleon-nucleon  potential\cite{AV18}. 
At
low values of $Q^2$, calculations\cite{Glock1,Glock2} including FSI and 
MEC, but
within a  non relativistic approach, are also shown.  At high 
values of $Q^2$, in the region close to the quasi-elastic peak, the quite 
reasonable
description of the data achieved
 by our PWIA (that contains relativistic effects) 
  confirms the physical expectation of the minor role played by FSI, when
the nucleon rapidly gets out. 
\begin{figure}
\label{q16}
\resizebox{4.1in}{!}{ \includegraphics{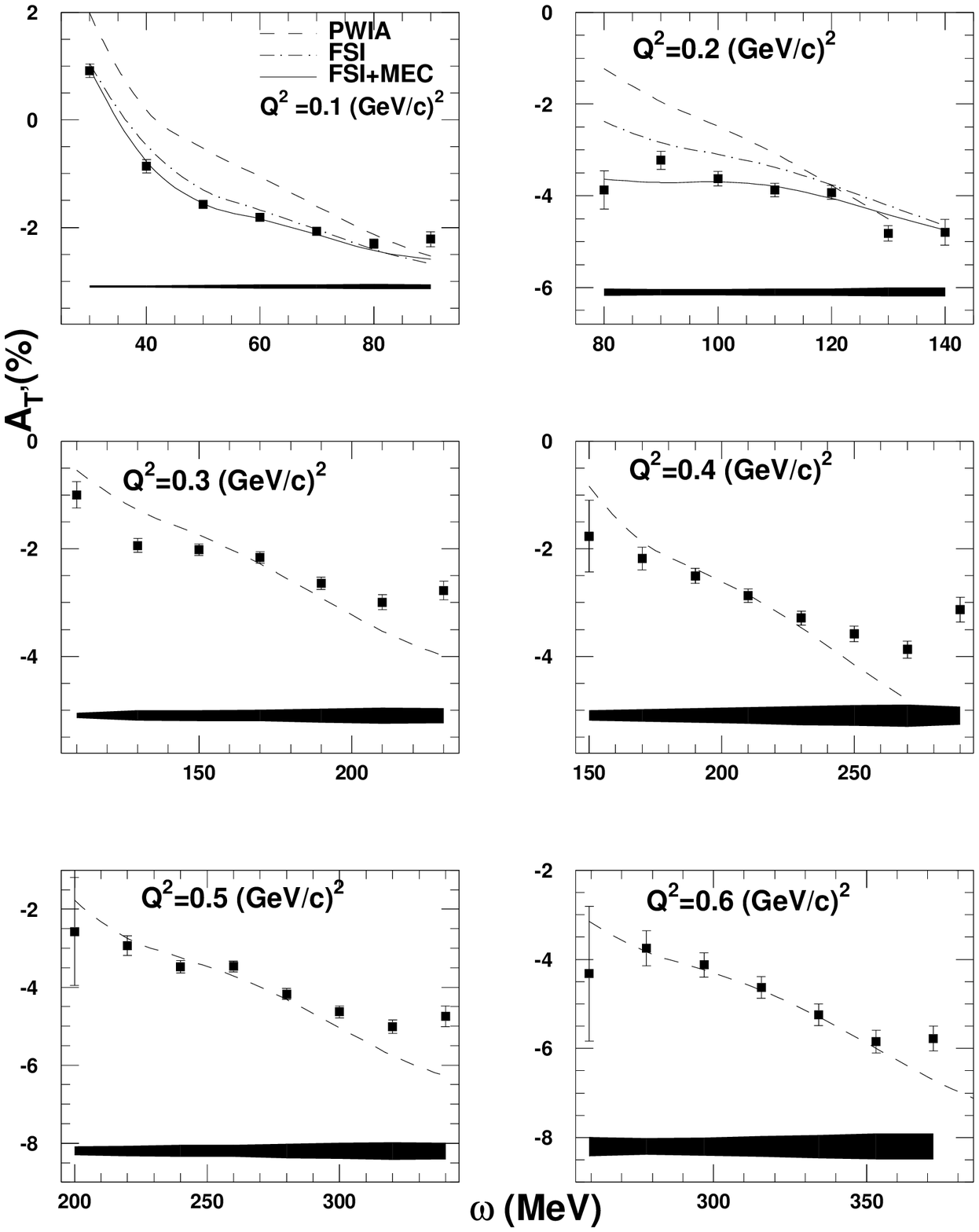} }

  \caption{ 
 The  asymmetry A vs  the energy transfer, $\omega$,
for different values of $Q^{2}$. Dashed
}
 {\footnotesize {lines: PWIA calculations within our approach   \cite{KPSV};
dash-dotted lines and solid lines: Faddeev calculations with FSI only  
  and with FSI +
MEC, respectively\cite{Glock1,Glock2}. Note that A is proportional to
 the transverese
$R_{T'}$ only close to the qe peak. 
  (After W. Xu et al.   \cite{Xu})}} 
  \end{figure}

\section{Three-nucleon scattering states} 
 
The fully-interacting, intrinsic wave function  for a three-nucleon
system in the continuous 
spectrum, 
can be decomposed   \cite{KRV}
 as follows
\be
\Phi^{jj_zTT_z \pi}=\Psi_A^{jj_zTT_z\pi}+\Psi_C^{jj_zTT_z\pi}
=\sum_{i=3} \left [\psi_A^{jj_zTT_z\pi}(i)+\psi_C^{jj_zTT_z\pi}(i)\right ]
 \label{wfsi} \ee
 where   $\Psi_A^{jj_zTT_z \pi}$ 
 is the solution of the Schr\"odinger equation in the
 asymptotic region, with  two well separated clusters,
  $\Psi_C^{jj_zTT_z \pi}$ describes the system when the three
 nucleons are close each other. The intrinsic energy is understood. 
 The functions $\psi_A^{jj_zTT_z\pi}(i)$ and 
 $\psi_C^{jj_zTT_z \pi}(i)$ are 
 Faddeev-like amplitudes, corresponding to the three permutations of the
 intrinsic coordinates ($ \equiv \{ \bf{r}_1,~\bf{r}_2,~\bf{r}_3 \}$). 
 
 The
 asymptotic component, 
  $\Psi_A^{jj_zTT_z \pi}$, can be recast in a different way, in order to
 emphasize its physical content. If 
 one considers the case of a N-d  scattering state, 
 for the sake of concreteness,  one has
 \be  
 \Psi_A^{LXjj_zTT_z \pi}= \Omega^R_{LXj}({\bf x}_1,
{\bf y}_1)+\nonu +
\sum_{L'X'}{}^j{\mathcal L}^{XX'}_{LL'}
\left [\imath ~
\Omega^R_{L'X'j}({\bf x}_1,{\bf y}_1)+\Omega^I_{L'X'j}({\bf x}_1,{\bf
y}_1)\right ]+  \nonu + \left [\psi_A^{LXjj_zTT_z}(2)+
 \psi_A^{LXjj_zTT_z}(3)\right ] \label{asy}\ee
 where $L$ is the relative orbital angular momentum of  N with respect to the deuteron,
 $X$ is the intermediate coupling of the spin of the nucleon with the total
 angular momentum of the deuteron, and $\{{\bf
 x}_1,~{\bf y}_1\}$ are the intrinsic coordinates  (${\bf x}_1={\bf r}_2-{\bf r}_3$ and $ 
 {\bf y}_1=  [{\bf r}_2+{\bf r}_3 -2{\bf
 r}_1  ]/\sqrt{3}$).
  
 In Eq. (\ref{asy}), $\Omega^{R(I)}_{LXj}$ represents  the regular 
 ("irregular", but properly regularized at small distances\cite{KRV}) solution describing
 the free scattering of a nucleon by an interacting pair (in this case 
  a deuteron); 
 the matrix ${\mathcal L}$ is given by\be 
 {\mathcal L}
 = {{\mathcal S}-1 \over 2 \imath}= -\pi {\mathcal T}  
 \label{sm}\ee 
 with ${\mathcal S}$ and ${\mathcal T}$  the S-matrix and the
 T-matrix, respectively.
Analogous expressions hold for $\psi_A^{LXjj_zTT_z}(2)$ and 
 $\psi_A^{LXjj_zTT_z}(3)$.
  In Eq. (\ref{asy}), 
  the first term produces the PWIA,
 i.e. contains an interacting spectator pair and a free particle; 
 the second term
 describes the rescattering between the interacting pair and the  asymptotically
 free particle;
the third term, $\left [\psi_A^{LXjj_zTT_z}(2)+
 \psi_A^{LXjj_zTT_z}(3)\right]$, takes care of the correct antisymmetrization of 
 $\Psi_A^{jj_zTT_z \pi}$.

The core component, $\Psi_C^{jj_zTT_z \pi}$, goes to zero 
 for large interparticle distances and energies
 below the    deuteron breakup threshold, while for  higher
 energies,  
 must reproduce a
 three outgoing particle state. In the  approach developed in Ref.
     \refcite{KRV}, 
  $\Psi_C^{jj_zTT_z \pi}$ is explicitly written as an expansion on a basis of
 Hyperspherical Harmonics Polynomials, with the inclusion of    
 pair-correlation functions, to be determined along with the  elements of the 
 S-matrix (see Eq. (\ref{sm})),   
 through a variational procedure (complex Kohn variational principle).
 For an illustration of the  behavior of the
 three-nucleon states in   coordinate space,  see, e.g., Ref. \refcite{KPS04}. 
 
\section{Including FSI}
The   interaction between the three nucleons in the final
state is taken into account in the evaluation of the matrix elements of the em
current as follows (note that  the current operator is approximated by 
a sum of one-body
operators, without two-body contributions)
\be 
\langle j'
,j'_z;T',T'_z,\pi',\epsilon'_{int};\beta';{\bf  q}|J^{\mu} _{IA}(0)|
{1 \over 2},j_z;{1 \over 2}, T_z,\pi_{b},\epsilon_{b}; {\bf 0 }\rangle  = 
\nonu = 3\sum_{\sigma_1}\sum_{\sigma'_1}\sum_{\sigma_2}\sum_{\sigma_3}
 \int d{\bf k}_1~ d{\bf k}_2 ~ \langle j'
,j'_z;T',T'_z,\pi',\epsilon'_{int};\beta
|{\bf k}'_1,{\bf k}'_2, \sigma'_1,\sigma_2,\sigma_3\rangle ~ \times
\nonu  \langle {\bf q }+ {\bf k}_1,\sigma'_1|J^{\mu}_{ 
1,free}(0)|{\bf k}_1, \sigma_1\rangle ~\langle {\bf k}_1,{\bf k}_2,\sigma_1,\sigma_2,\sigma_3|
{1 \over 2},j_z;{1 \over 2}, T_z,\pi_{b},\epsilon_{b} \rangle
\label{FSI1} 
\ee
where $\langle {\bf k}_1,{\bf k}_2,\sigma_1,\sigma_2,\sigma_3|
j,j_z;T, T_z,\pi,\epsilon  \rangle$ indicates the intrinsic three-body wave 
function, related to  the state containing the center 
of mass motion   by the following simplified
expression
\be
\langle {\bf p}_1,{\bf p}_2,{\bf p}_3, \sigma_1,\sigma_2,\sigma_3|
j,j_z;T, T_z,\pi,\epsilon;{\bf P}  \rangle  =
~\delta({\bf p}_1+{\bf p}_2+{\bf p}_3-{\bf P})~\times \nonu
\langle {\bf k}_1,{\bf k}_2,\sigma_1,\sigma_2,\sigma_3|
j,j_z;T, T_z,\pi,\epsilon  \rangle
\label{cm}
\ee
where ${\bf k}_i$ is the spatial part of the four-momentum $k^{\mu}_i \equiv \{\sqrt{m^2+|{\bf k}_i|^2},{\bf k}_i \} =B^{\mu}_{\nu} ({\bf
P}/M)p^{\mu}_i$, with $B^{\mu}_{\nu}$ the  boost transformation. 
 The expression in Eq. (\ref{cm}) is an
approximate one, since the Wigner functions corresponding to the boost
transformations, as well as other kinematical factors, are
dropped out  in this
 introductory presentation of our approach.
 
\begin{figure}
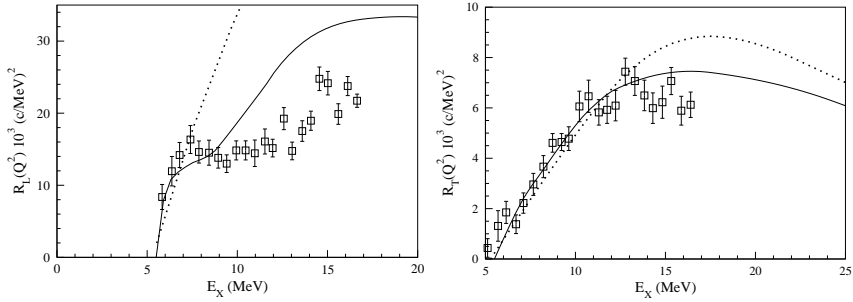

\parbox{2.2in}{\resizebox{2.2in}{!}{ \includegraphics{RL_882ws.eps}}}
\parbox{2.2in}{\resizebox{2.2in}{!}{ \includegraphics{RT_882ws.eps} }}

\caption{ Unpolarized response functions, $R_L$ and $R_T$, vs the 
the missing energy 
$E_X=\sqrt{(\omega+M_{He})^2-|\vec{q}|^2}-M_{He}\simeq
B_3+\epsilon'_{int}$ 
for  $|\vec{q}|\sim 175~MeV$. Solid line: preliminary}
{\footnotesize{results with FSI; 
  dotted line: PWIA.
 Experimental
data from  Ref.  \refcite{Retz}.}}
\end{figure}

\begin{figure}
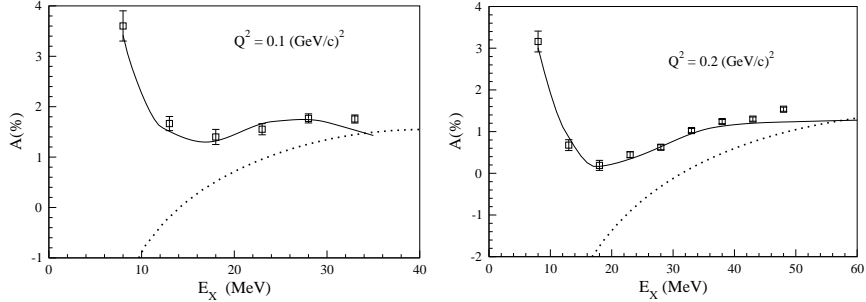

\parbox{2.2in}{\resizebox{2.2in}{!}{ \includegraphics{AT_x1ws.eps}}}$~~$
\parbox{2.2in}{\resizebox{2.2in}{!}{ \includegraphics{AT_x2ws.eps} }}

\caption{ The asymmetry A for $Q^2=0.1~(GeV/c)^2$ (left) and   
$Q^2=0.2~(GeV/c)^2$ (right) vs the missing energy, $E_{\rm X}$, at low
energy transfer, as in Fig. 2. Solid line: preliminary results with FSI; 
  dotted line: PWIA. Experimental data from F. Xiong et}{\footnotesize{  
 al.  \cite{Xiong}. }}
\end{figure}
Our  preliminary calculations based on the matrix elements
of the em current of $^3$He, Eq. (\ref{FSI1}), both for the unpolarized ($R_L$
and $R_T$) responses and the polarized asymmetry
 are presented in Figs. 2,3 and 4. FSI 
is taken
into account only in the two-body break-up channel 
 (where
 a fully interacting three-nucleon state  behaves  
asymptotically  like a $pd$ system). In particular we have included FSI up to 
 $j'\le 5/2$ for a CM energy of the three-nucleon system $E^{fin}_{CM} \le 10
 ~MeV$, up to $j'~ \leq~
7/2$ for $10~MeV< E^{fin}_{CM} \le 30~ MeV$, and  up to $j'~ \leq~
11/2$ for $30~MeV< E^{fin}_{CM} < 100 MeV$. For all other waves we used the PWIA
 (up to $j'~ \leq~
17/2$ for $Q^2=0.1 ~(GeV/c)^2$ and up to $j'~ \leq~
19/2$ for $Q^2=0.2 ~(GeV/c)^2$). In the three-body breakup channel we always
adopted PWIA. The AV18 nucleon-nucleon interaction \cite{AV18} has been used, 
without Coulomb effects nor three-body forces
 (that will be included in the future). 
The strong effect of FSI is fully confirmed and
 we obtain  
results in qualitative agreement with the ones by the Bochum 
group\cite{Glock1,Glock2}, though in our calculations FSI are
presently taken into account only in the
two-body break-up channel.

\begin{figure}
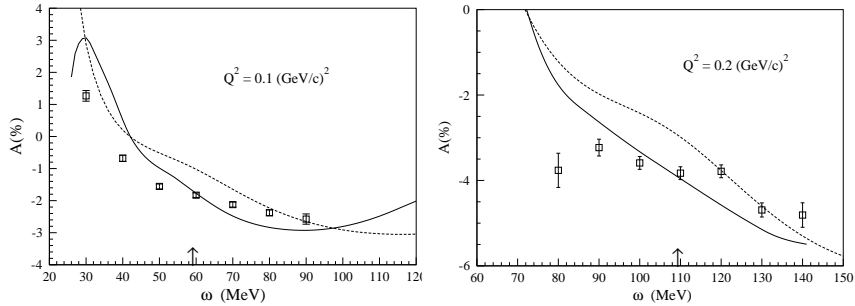

\parbox{2.2in}{\resizebox{2.2in}{!}{ \includegraphics{AT_01ws.eps}}}
\parbox{2.2in}{\resizebox{2.2in}{!}{ \includegraphics{AT_02ws.eps} }}

\caption{The asymmetry A  for $Q^2=0.1~(GeV/c)^2$ (left) and 
$Q^2=0.2~(GeV/c)^2$ (right) vs the
energy transfer, $\omega$, in the region of the quasi-elastic
peak. Solid line: preliminary results with FSI; 
  dotted line: PWIA. Experimental data from W. Xu et al.}
{\footnotesize{ 
   \cite{Xu,Xu2}, and private communication.
  Arrows indicate the qe peak, where the asymmetry A is proportional to R$_{T'}$}}
\end{figure}

\section{Summary and Perspectives}
Recently  the inclusive scattering of polarized electrons by a polarized
$^3$He target have been measured at TJLAB, both in the region of the quasi
elastic peak and in the low-$\omega$ wing\cite{Xu,Xu2}. The direct comparison of 
calculations corresponding to different theoretical approaches
with these experimental results 
allows one to better understand the model dependence in the extraction of the
neutron em properties, like the magnetic form factor $G_M^n$. Indeed for  a satisfactory interpretation
 of the data  one needs 
 accurate theoretical calculations that include effects beyond the
PWIA, such as i) FSI, ii) MEC, iii) relativistic effects and
iv) contributions from the explicit presence of the $\Delta$ excitation in the
ground state of $^3$He. 
In  our  calculations we have adopted both relativistic kinematics and a relativistic
electron-nucleon cross section,  and we have taken into account
exactly the FSI in the two-body break-up channel, by using the 
 three-nucleon wave functions  obtained 
 by the
Pisa group\cite{KRV} within a variational approach for both the bound and
the excited states.

The development of our approach will follow two distinct paths: i) 
a better treatment of the relativistic effects within the so called Relativistic
Hamiltonian Dynamics (see, e.g., B.D. Keister and W. Polyzou  \cite{KP}), that 
 allows  a Poincar\'e covariant  description of an interacting system
with a fixed number of particles; ii) the inclusion of FSI in the  three-body break-up
channels, together with Coulomb effects and three-body forces.
 
Our present technology, based on the overlaps appearing in Eq. (\ref{FSI1}) 
can be easily generalized to calculate the pion electroproduction from a
polarized  $^3$He target.
%
%

\end{document}